\documentclass[conference]{IEEEtran}
\IEEEoverridecommandlockouts
\usepackage{cite}

\usepackage{amsmath,amssymb,amsfonts}

\usepackage{bbm}

\usepackage{mathrsfs}

\usepackage{dsfont}

\usepackage{array}

\usepackage{epstopdf}

\usepackage{setspace}

\usepackage{graphics}

\usepackage{graphicx}

\usepackage{textcomp}

\usepackage{xcolor}

\usepackage{multirow}

\usepackage{xcolor}

\usepackage{stfloats}

\usepackage{subfigure}

\usepackage{colortbl}

\usepackage{pdflscape}

\usepackage{dsfont}

\usepackage{accents,amsmath,amssymb,amsfonts,amsthm}
\usepackage{algorithm, algpseudocode}
\usepackage[square, comma, sort&compress, numbers]{natbib}

\IEEEoverridecommandlockouts
\theoremstyle{definition}
\newtheorem{theorem}{Theorem}

\newtheorem{corollary}{Corollary}
\newtheorem{remark}{Remark}
\newtheorem{proposition}{Proposition}
\newtheorem{case}{Case}
\allowdisplaybreaks

\usepackage{colortbl}
\definecolor{gray1}{gray}{.9}
\definecolor{pink1}{rgb}{.99,.91,.95}
\definecolor{cyan1}{cmyk}{.3,0,0,0}

\makeatletter
\def\widebar{\accentset{{\cc@style\underline{\mskip10mu}}}}
\def\Widebar{\accentset{{\cc@style\underline{\mskip15mu}}}}
\makeatother

\newcommand{\red}[1]{\textcolor{red}{#1}}

\newcommand{\gray}[1]{\textcolor{gray}{#1}}

\def\BibTeX{{\rm B\kern-.05em{\sc i\kern-.025em b}\kern-.08em
    T\kern-.1667em\lower.7ex\hbox{E}\kern-.125emX}}
\begin{document}



\title{Queue-Aware Variable-Length Coding for\\ Ultra Reliable Low Latency Communications \\with Random Arrival}

\author{\IEEEauthorblockN{Xiaoyu~Zhao,~Wei~Chen,~\textit{Senior~Member,~IEEE}}
\IEEEauthorblockA{Beijing National Research Center for Information Science and Technology\\
	Department of Electronic Engineering, Tsinghua University, Beijing, 100084, CHINA \\
Email: xy-zhao16@mails.tsinghua.edu.cn, wchen@tsinghua.edu.cn}
}
\maketitle

\begin{abstract}
	With the phenomenal growth of the Internet of Things (IoT), Ultra Reliable Low Latency Communications (URLLC) has potentially been the enabler to guarantee the stringent requirements on latency and reliability.
	However, how to achieve the low latency and ultra reliability with the random arrival remains open.
	In this paper, a queue-aware variable-length channel coding is presented over the single URLLC user link, in which the finite blocklength of channel coding is determined based on the random arrival.
	More particularly, a cross-layer approach is proposed for the URLLC user to establish the optimal tradeoff between the latency and the power consumption.
	With a probabilistic coding framework presented, the cross-layer variable-length coding can be characterized based on a Markov chain.
	In this way, the optimal delay-power tradeoff is given by formulating an equivalent Linear Programming (LP).
	By solving this LP, the delay-optimal variable-length coding can be presented based on a threshold-structure on the queue length.%
\end{abstract}%

\section{Introduction}
As one of the major concerns in the Fifth Generation (5G) wireless system, URLLC is mainly expected to support ultra-low latency transmission with ultra-high reliability satisfied.
In particular, URLLC has attracted significant attention in the IoT applications, such as the factory automation, smart grids, and Intelligent Transport System (ITS) \cite{URLLC_IoT}.
To transmit critical information in those mission-critical communications, URLLC considers the requirements of reliability and latency based on the $10^{-5}$ packet loss probability and $1$ms end-to-end delay \cite{TS_req}.



To satisfy the stringent requirements in URLLC, the finite blocklength channel coding can be employed in URLLC based on a cross-layer design approach.
With the short frame structure utilized in URLLC \cite{short_frame,TTI_short}, the finite blocklength channel coding is presented to satisfy the low latency and ultra reliability.
For the finite blocklength, the tradeoff between latency, reliability, rate has been specifically developed based on the nonasymptotic information theory in \cite{capacity_short}.
To jointly consider the end-to-end latency and ultra reliability in URLLC, a cross-layer design shall be employed to optimize the finite blocklength coding with the random arrival.

As the efficient way to satisfy Quality of Service (QoS), the cross-layer designs have been widely investigated in the resource allocation, scheduling policy, and access control.
The pioneering work on the cross-layer scheduling was proposed by Collins and Cruz in \cite{collins1999transmission}, where the average delay and power are studied over a single user link.
In \cite{berry2002communication}, the tradeoff between the latency and power consumption was presented by Berry and Gallager.
Based on the probabilistic scheduling policies, we focused on the optimal delay-power tradeoff under the different \mbox{scenarios} \cite{chen2017delaytcom,meng2018tcom}.
The cross-layer designs have been also investigated in URLLC.
With the packet dropping policy, the power consumption was minimized in \cite{URLLC_cross_layer_2} under the reliability constraint, where the violation probability of the maximal latency is included.
Moreover, the violation probabilities of delay and peak-age of information were given for URLLC in \cite{URLLC_cross_layer_1}.

In this paper, a cross-layer design of the variable-length coding is presented for the signal URLLC user link.
By this means, the optimal delay-power tradeoff of the URLLC user is formulated to meet the latency and reliability requirements.
With the short time duration in the URLLC user, the finite blocklength variable-length coding is employed to transmit the random arrival data traffic.
Under the aware queue length of data packets, the cross-layer variable-length coding determines the channel coding blocklength in terms of its probability.

With the queue-aware variable-length coding policy, the requirements of reliability and end-to-end latency can be jointly considered in URLLC.
In particular, the reliability requirement is satisfied based on the power allocation for the variable-length coding.
Meanwhile, the optimal average delay is presented for the URLLC user based on the optimal delay-power tradeoff, where the average delay is minimized under the average power constraint.
By formulating the equivalent LP problem, the optimal delay-power tradeoff is particularly attained by the delay-optimal variable-length coding.
\mbox{Moreover,} the optimal variable-length coding is presented based on the threshold-based structure on the queue length.
With the aware queue length, the delay-optimal blocklength of the channel coding can be presented under the power constraint.


\section{System Model}

\begin{figure}[t]
	\centering
	\includegraphics[width=1\columnwidth]{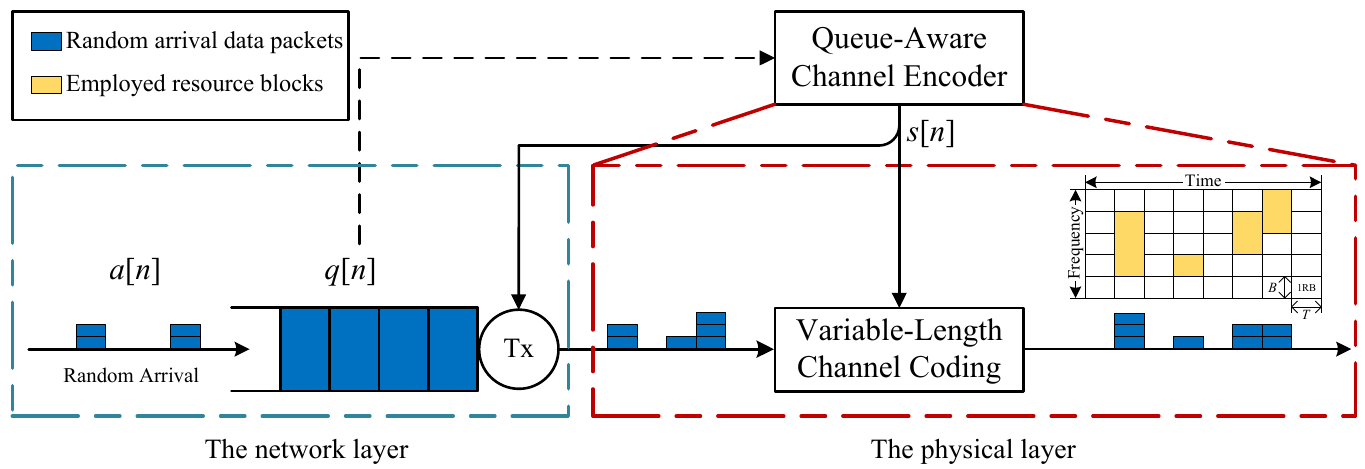}
	\vspace{-8mm}
	\caption{System Model.}
	\label{fig_model}
	\vspace{-6mm}
\end{figure}

As shown in Fig. \ref{fig_model}, the queue-aware variable-length coding is investigated for the single user link in URLLC.
Under the cross-layer design, the blocklength of coding in the physical layer is determined by queue length in the network layer, where the packets are randomly arrive and temporarily backlog.
We next present the variable-length coding in the physical layer and the queue model in the network layer.

\subsection{The Variable-Length Coding in the Physical Layer}

In the physical layer, the data packets are transmitted by the variable-length coding over the Addition White Gaussian Noise (AWGN) channel, in which we denote by ${N_0}$ the single-sided noise spectral density.
Meanwhile, the time-frequency resource is divided into a set of 
orthogonal Resource Blocks (RB), all of which have the same time duration $T$ and frequency bandwidth $B$.
In this way, the time is divided into the timeslots, where the time duration is equal to $T$.

In the variable-length coding, the blocklength is next given based on the number of the packets that are jointly encoded in each timeslot.
More specifically, we denote by $s[n]$ the number of the packets transmitted at the $n$th timeslot.
The $s[n]$ packets are then encoded on the $s[n]$ orthogonal RBs, which are in the same timeslot.
By this means, the blocklength for the variable-length channel coding is determined as $TBs[n]$ for the $n$th timeslot.
Considering the finite power at the encoder, we have that $s[n]$ belongs to set $\{0,1,\cdots,S\}$ for each timeslot.

With the short time duration employed in URLLC \cite{TTI_short}, the blocklength of the variable-length coding is far from infinite in each timeslot.
Based on the normal approximation in \cite{capacity_short}, the number of packets $s[n]$ that are transmitted by the variable-length coding can be approximately given by the power consumption $P[n]$ in the $n$th timeslot.
With the Single-to-Noise Ratio (SNR) $\gamma[n]$ defined as $\frac{P[n]}{N_0Bs[n]}$, the tight approximation is particularly shown as
\vspace{-1mm}
	\begin{equation}\label{R_C_short}
	\!\!s[n]\!\approx\!\frac{T\!Bs[n]}{L\ln{2}}\!\!\left(\!\ln{\!\left(1\!+\!\gamma[n]\right)}\!-\!\!\sqrt{\!\frac{\gamma[n](2\!+\!\gamma[n])}{T\!Bs[n](1\!+\!\gamma[n])^2}}Q^{-1}\!(\epsilon)\!\right)\!\!,
	\vspace{-1mm}
	\end{equation}
where $Q^{-1}(\cdot)$ is the inverse of the Gaussian Q-function, and the error probability and the number of bits in one packet are defined as $\epsilon$ and $L$, respectively. 
Meanwhile, the approximation holds up to a term of $\mathcal{O}(\frac{\log{TBs[n]}}{TBs[n]})$.

Moreover, we can determine the power consumption $P[n]$ for the variable length coding, under which the $s[n]$ packets are successfully transmitted within the error probability $\epsilon$. 
In particular, $P[n]$ is given by the function $P(s)$, i.e., $P[n]=P(s[n])$, which is formulated according to Eq. (\ref{R_C_short}).
With the reliability satisfied by $P[n]$, the queue-aware variable-length coding is then presented based on a cross-layer design.

\subsection{The Queue Model in the Network Layer}

In the network layer, the random arrival packets will be temporarily stored at the URLLC user's buffer before the packets are transmitted by the variable-length channel coding.
More specifically, the random data arrival at the beginning of timeslots follows the Bernoulli process with the probability \mbox{as $\alpha$.}
For each arrival, $A$ packets will arrive at the user's buffer, where we have $A\le{}S$.
We denote by $a[n]$ the number of the packets arrive at the $n$th timeslot.
In this way, we have $\Pr\{a[n]=A\}=\alpha$, $\Pr\{a[n]=0\}=1-\alpha$ in each timeslot.

The arrival data packets will be then backlogged at the user's buffer, in which the buffer size is $Q$.
Meanwhile, the buffer state will dynamically evolve in terms of the buffer's queue length, which is denoted by $q[n]$ and belongs to set $\{0,1,\cdots,Q\}$.
The dynamical evolution of the queue length $q[n]$ is presented as
\vspace{-1mm}
\begin{equation}
	q[n+1]=\min\{(q[n]-s[n])^++a[n+1],Q\},
	\vspace{-1mm}
\end{equation}
where we have $(x)^+=\max\{x,0\}$.

By this means, the queue-aware variable-length coding is finally formulated to present the blocklength $TBs[n]$ for the aware queue length $q[n]$.
Since the blocklength is determined by $s[n]$, we particularly present the variable-length coding policy by defining the probabilities $f_{q,s}$ as
\vspace{-1mm}
\begin{equation}
	f_{q,s}=\Pr\{s[n]=s|q[n]=q\},
	\vspace{-1mm}
\end{equation}
where we have $\sum_{s=0}^Sf_{q,s}=1$.
Therefore, the variable-length coding policy $\boldsymbol{F}$ is \mbox{given by $\{f_{q,s}:0\le{}q\le{}Q,0\le{}s\le{}S\}$.}

To avoid the underflow and overflow, we have that $s[n]$ shall satisfy $0\le{}q[n]-s[n]\le{}Q-A$.
In other words, the obtainable $s[n]$ in each slot is restricted by $s_{q[n]}^{\min}\le{}s[n]\le{}s_{q[n]}^{\max}$, where $s_{q}^{\min}$ and $s_{q}^{\max}$ are defined as  $\max\{0,q\}$ and $\min\{S,q-Q+A\}$, respectively. 
We then have $f_{q,s}=0$ if $s>s_{q}^{\max}$ or $s<s_{q}^{\min}$ for all the variable-length coding policies.
Based on the cross-layer design, the low latency and ultra reliability can be then jointly considered for the URLLC user.

\begin{figure*}[t]
	\centering
	\includegraphics[width=2\columnwidth]{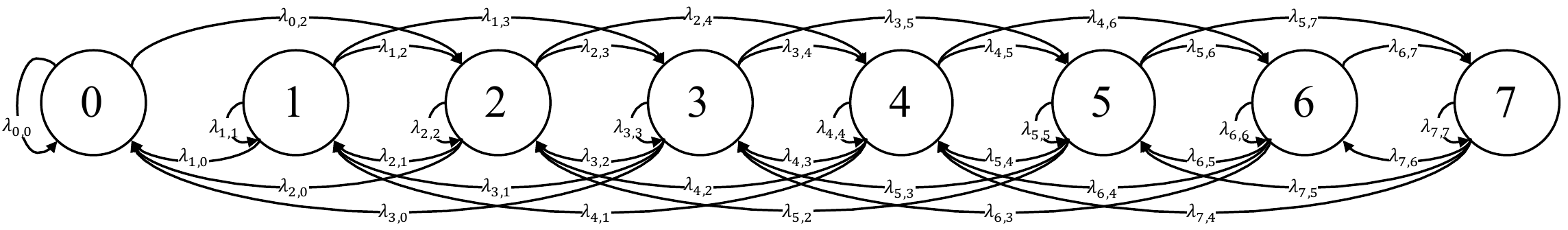}
	\vspace{-3mm}
	\caption{The transition diagram given by the Markov chain with $Q=7$, $A=2$, and $S=3$: the arrows present the transition relationships among the different queue length, and the corresponding transition probabilities are presented.}
	\label{transition_diagram}
	\vspace{-4mm}
\end{figure*}

\section{The Optimal Delay-Power Tradeoff under the Queue-Aware variable-length coding}

To satisfy the low latency and ultra reliability requirements in URLLC, the optimal delay-power tradeoff is formulated under the cross-layer variable-length coding.
Based on the queue-aware coding, the average delay and power consumption are first presented for the URLLC user.
With the average power constraint given, the average delay is then minimized by using the equivalent LP problem.
Moreover, the optimal delay-power tradeoff curve is formulated to presenting the optimal average delay with the varying power constraint.

\subsection{The Markov Chain under the Variable-Length Coding}

First, the long term average delay and power consumption are rigorously presented under the queue-aware variable-length coding.
To this end, the variable-length coding is formulated as the Markov chain with the queue length $q[n]$ as the state.
\mbox{In particular,} the Markov chain under the variable-length coding policy $\boldsymbol{F}$ can be presented by using the transition probability $\lambda_{i,j}$.
Considering the probability $\lambda_{i,j}$ is defined as $\Pr\{q[n+1]=j|q[n]=i\}$, we have
\vspace{-2mm}
\begin{equation}\label{Pr_transition}
	\lambda_{i,j}=\sum_{s=s_q^{\min}}^{s_q^{\max}}
				  \alpha{}f_{i,s}\mathds{1}_{\{i-s+A=j\}}+(1-\alpha){}f_{i,s}\mathds{1}_{\{i-s=j\}}.
\vspace{-2mm}
\end{equation}
In this way, the transitions among the different queue lengths can be present by the transition diagram as shown in \mbox{Fig. \ref{transition_diagram}}.

Meanwhile, the steady-state probability for queue length can be expressed based on the balance equations of the Markov chain.
We denote by $\pi_q$ the steady-state probability for queue length $q$.
Then, the balance equations of $\pi_q$ are presented as%
\vspace{-1mm}
\begin{equation}\label{balance_eq}
	\begin{cases}
		\sum_{i=\max\{0,q-A\}}^{\min\{Q,q+S\}}\pi_{i}\lambda_{i,q}=\pi_{q},&\forall~q=0,1,\cdots,Q,\\
		\sum_{q=0}^{Q}\pi_{q}=1.
	\end{cases}
	\vspace{-1mm}
\end{equation}
With the steady-state probability distribution $\boldsymbol{\pi}$ presented as vector $[\pi_0,\cdots,\pi_Q]^T$, the balance equations in Eq. (\ref{balance_eq}) can be expressed as
	\vspace{-1mm}
\begin{equation}
	\begin{cases}
		\boldsymbol{\Lambda}_{\boldsymbol{F}}\boldsymbol{\pi}=\boldsymbol{\pi},\\
		\boldsymbol{1}^T\boldsymbol{\pi}=1,
	\end{cases}
	\vspace{-1mm}
\end{equation}
where we formulate the matrix $\boldsymbol{\Lambda}_{\boldsymbol{F}}$ by presenting $\lambda_{i,j}$ as the element at the $(j+1)$-th row and $(i+1)$-th column, and denote by $\boldsymbol{1}$ the vector with all the \mbox{elements as $1$.}

Based on the steady-state probabilities $\pi_q$, the average delay and power consumption are next presented for the URLLC user.
According to Little's Law, we can formulate the average delay $D_{\boldsymbol{F}}$ under variable-length coding policy $\boldsymbol{F}$ as
\vspace{-1.5mm}
\begin{equation}
	D_{\boldsymbol{F}}=\frac{1}{A\alpha}\sum_{q=0}^{Q}q\pi_q,
\vspace{-1.5mm}
\end{equation}
where the arrival rate of the data packets is equal to $A\alpha$.
Similarly, the average power consumption is formulated as
\vspace{-1.5mm}
\begin{equation}
P_{\boldsymbol{F}}=\sum_{q=0}^{Q}\sum_{s=s_q^{\min}}^{s_q^{\max}}P(s)f_{q,s}\pi_q,
\vspace{-1.5mm}
\end{equation}
where we recall function $P(s)$ presents the power consumption of the variable-length coding.
In particular, the function $P(s)$ can be given by the following theorem. 
\vspace{-1mm}
\begin{theorem}\label{th_P_s}
	With the error probability $\epsilon$ given, the function $P(s)$ is approximately expressed as the parametric equation
	\vspace{-1mm}
	\begin{equation}\label{P_s_short}
	\begin{cases}
	s~\approx\frac{\gamma(\gamma+2)}{\left(\ln{(1+\gamma)}-\frac{L}{BT}\ln{2}\right)^2(1+\gamma)^2}\frac{\left(Q^{-1}(\epsilon)\right)^2}{BT},\\
	P\approx\frac{\gamma^2(\gamma+2)}{\left(\ln{(1+\gamma)}-\frac{L}{BT}\ln{2}\right)^2(1+\gamma)^2}\frac{\left(Q^{-1}(\epsilon)\right)^2N_0}{T},
	\end{cases}
	\vspace{-1mm}
	\end{equation}
	where the SNR $\gamma$ satisfies $\gamma>2^{\frac{L}{BT}}-1$, and the approximation error vanishes as $BT\rightarrow+\infty$.
\end{theorem}
	\vspace{-2mm}
\begin{IEEEproof}
	Our proof starts with the observation that the relationship of $s[n]$ and $\gamma[n]$ in Eq. (\ref{R_C_short}), by which $s$ is immediately presented according to Eq. (\ref{P_s_short}).
	Since $P$ is equal to $N_0Bs\gamma$, we have $P$ is also presented by Eq. (\ref{P_s_short}).
	To make the expression of $P(s)$ sensible, we have that $\ln(1+\gamma)>\frac{L}{TB}\ln2$, i.e., $\gamma>2^{\frac{L}{TB}}-1$, which completes the proof.	
\end{IEEEproof}

According to the function $P(s)$ expressed in Eq. (\ref{P_s_short}), the following corollary can be straightforwardly presented.
\vspace{-2mm}
\begin{corollary}\label{co_power_eff}
	The function $P(s)$ satisfies that
	\begin{enumerate}
		\item The function $P(s)$ is concave on set $[s^\ast,+\infty)$, where $s^\ast$ is the only inflection point on $P(s)$;
		\item The power consumption for each data packet, i.e., $\frac{P}{s}$, decreases with the number of data packets $s$.
	\end{enumerate}
\end{corollary}%
\vspace{-2mm}
\begin{remark}
Considering $s^\ast$ is much less than $1$, we nearly consider sequence $\{P(s):s\!=\!0,\cdots,S\}$ is concave on $s$.
\end{remark}
\vspace{-2mm}

\subsection{The equivalent Linear Programming problem}

Then, the optimal delay-power tradeoff is presented for the URLLC user under the cross-layer variable-length coding.
More specifically, the cross-layer optimization problem is formulated to minimize the average delay with the average power constraint.
In other words, we have
\begin{subequations}
\begin{align}
	\min_{\{f_{q,s},\pi_q\}}&D_{\boldsymbol{F}}\\
	\rm{s.t.~}&P_{\boldsymbol{F}}\le{}P_{\text{th}}\\
			  &\boldsymbol{\Lambda}_{\boldsymbol{F}}\boldsymbol{\pi}=\boldsymbol{\pi}\\
			  &\boldsymbol{1}^T\boldsymbol{\pi}=1\\
			  &\!\!\!\begin{array}{l}
			  \sum_{s=s_q^{\min}}^{s_q^{\max}}f_{q,s}=1,
			  \end{array}~~\forall~q\\
			  &\pi_q\ge{}0,~f_{q,s}\ge{}0,~~\forall~q,~s,
\end{align}\label{D_P_tradeoff}%
\end{subequations}
where we denote by $P_{\text{th}}$ the average power constraint.

Moreover, the equivalent LP problem is next \mbox{formulated} to attain the optimal delay-power tradeoff.
In particular, the equivalent LP problem is presented in the following theorem.%
\begin{theorem}
	\label{th_LP}
	The problem (\ref{D_P_tradeoff}) is equivalent to the following linear programming problem.
	\begin{subequations}
		\begin{align}
		\min_{\{x_{q,s}\}}&
		\begin{array}{l}
			\!\!\frac{1}{A\alpha}\sum_{q=0}^{Q}\sum_{s=s_q^{\min}}^{s_q^{\max}}qx_{q,s}
		\end{array}\label{LP_D}\\
		~\rm{s.t.~}&\begin{array}{l}
			\!\!\sum_{q=0}^{Q}\sum_{s=s_q^{\min}}^{s_q^{\max}}P(s)x_{q,s}
		\end{array}\le{}P_{\rm{th}}\label{LP_c_P}\\
		&\begin{array}{l}
		\!\!\sum_{i=\max\{0,q-A\}}^{\min\{Q,q+S\}}\sum_{s=s_i^{\min}}^{s_i^{\max}}\alpha{}x_{i,s}\mathds{1}_{\{i-s+A=q\}}\\
		\!\!+(1-\alpha){}x_{i,s}\mathds{1}_{\{i-s=q\}}\!=\!\sum_{s=s_{q}^{\min}}^{s_{q}^{\max}}x_{q,s}, ~\forall~q
		\end{array}\label{LP_c_1}\\
		&\begin{array}{l}
		\!\!\sum_{q=0}^{Q}\sum_{s=s_q^{\min}}^{s_q^{\max}}x_{q,s}
		\end{array}=1 \label{LP_c_2}\\
		&~~\!\!x_{q,s}\ge{}0,~\forall~q,~s \label{LP_c_3}\\
		&~~\!\!x_{q,s}={}0,~\forall~q-s<0~{\rm{or}}~q-s>Q-A \label{LP_c_4}.
		\end{align}\label{D_P_LP}%
	\end{subequations}
\end{theorem}%
\vspace{-14pt}
\begin{IEEEproof}
The main idea of the proof is to construct the bijective map between the feasible solutions for problems (\ref{D_P_tradeoff}) and (\ref{D_P_LP}).
The details of the map construction can be found in our previous work presented in \cite[Section IV-D]{chen2017delaytcom}.
\end{IEEEproof}

In the equivalent LP problem, we define the optimization variable $x_{q,s}$ as $f_{q,s}\pi_q$, which indicates the frequency of $q[n]=q$ and $s[n]=s$ over the long term.
With the optimization variable $x_{q,s}$ employed, problem (\ref{D_P_tradeoff}) can be immediately rewritten as LP problem (\ref{D_P_LP}).

\vspace{-1mm}
\subsection{The Optimal Delay-Power Tradeoff Curve}

The optimal delay-power tradeoff curve is finally formulated for the URLLC user to present the minimized average delays under the different power constraints.
To this end, we present the equivalent LP problem on \mbox{the power-delay plane}, in which the average power-delay pairs are generated by the variable-length coding policies.
More specifically, we have
\begin{subequations}
	\begin{align}
	\min_{(\widebar{P},\widebar{D})\in\mathcal{R}}&\widebar{D}\\
	~\rm{s.t.~}&\widebar{P}\le{}P_{\rm{th}},
	\end{align}%
\end{subequations}%
where set $\mathcal{R}$ consists of all the obtainable average power-delay pairs under all the variable-length coding policies.

\begin{figure}[t]
	\centering
	\includegraphics[width=0.8\columnwidth,height=0.65\columnwidth]{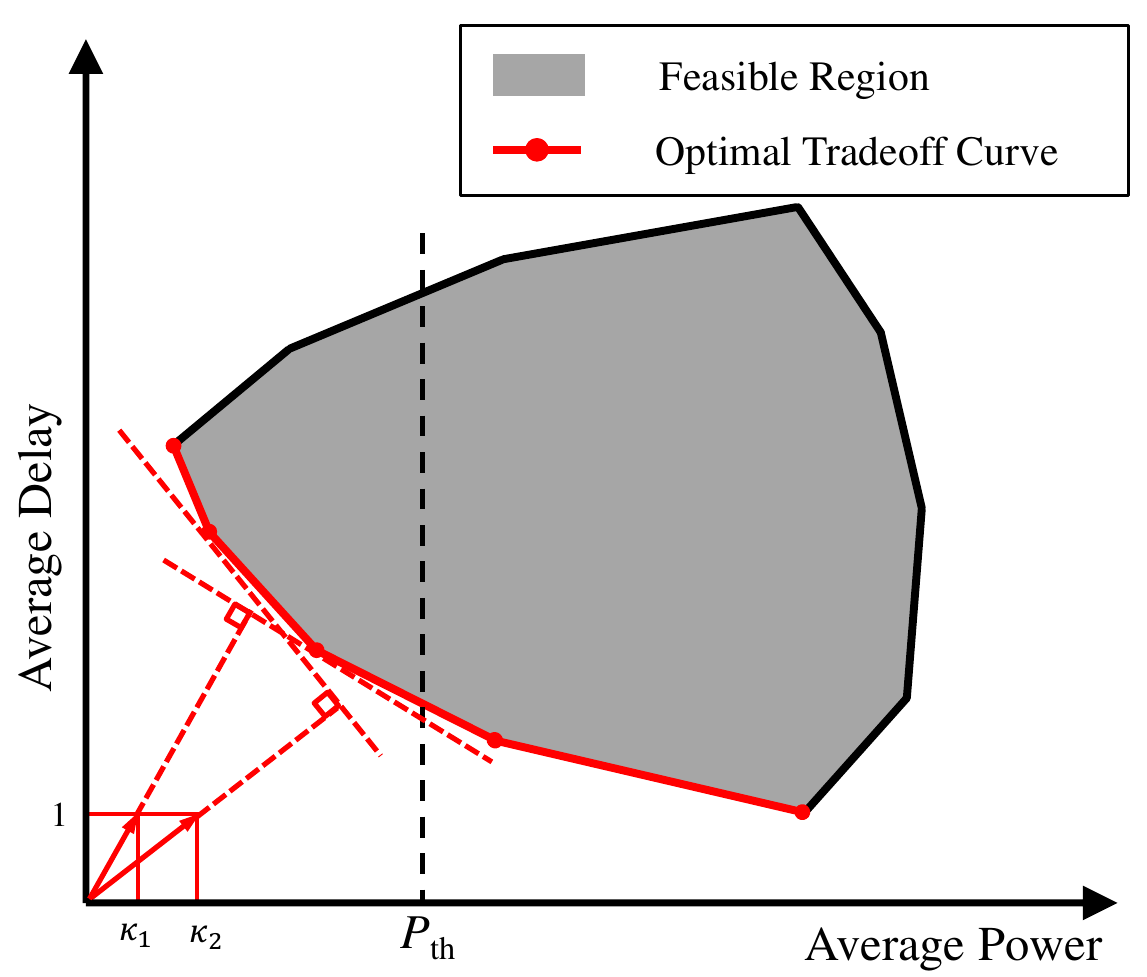}
	\vspace{-4mm}
	\caption{The sketch of the optimal delay-power tradeoff curve.}
	\label{fig_cure}
	\vspace{-6mm}
\end{figure}

Moreover, the set $\mathcal{R}$ can be shown as the polyhedron on the power-delay plane based on the linear expressions of the average delay and power in Eqs. (\ref{LP_D}) and (\ref{LP_c_P}).
In particular, the average power-delay pairs can be generated as the projection of the obtainable $\{x_{q,s}\}$ on the power-delay plane.
Considering all the obtainable $\{x_{q,s}\}$ are given by the linear constraints in Eqs. (\ref{LP_c_1})-(\ref{LP_c_4}), the set consisting of all $\{x_{q,s}\}$ is then shown as the closed bounded polyhedron.
As the projection on the power-delay plane, the set $\mathcal{R}$ is formulated as the polyhedron with the sketch presented as Fig. \ref{fig_cure}.

With set $\mathcal{R}$ given, the optimal delay-power tradeoff curve can be then formulated in the delay-power plane.
As demonstrated as Fig. \ref{fig_cure}, the average delay is minimized over set $\mathcal{R}$ with the power constraint satisfied.
Meanwhile, the geometric properties of the optimal delay-power tradeoff curve are presented as the following theorem.
\vspace{-1mm}
\begin{theorem}\label{th_curve}
	The optimal delay-power tradeoff curve is piecewise linear, decreasing, and convex.
\end{theorem}
\vspace{-1mm}
\begin{IEEEproof}
	The theorem is straightforwardly proofed based on the method presented in \cite[Corollary 3]{chen2017delaytcom}.
\end{IEEEproof}

Therefore, the optimal average delay can be obtained for the URLLC user under the cross-layer variable-length coding.

\section{The Threshold-based Structure in The Optimal Variable-Length Coding Policy}
\vspace{-1mm}

In this section, the optimal variable-length coding policy is presented by exploiting the threshold-based structure on the queue length.
Based on the equivalent LP problem, we first present the optimal variable-length coding, by which the optimal delay-power tradeoff is shown under the different initial queue lengths.
Then, the optimal blocklength of the variable-length coding is presented as the maximal or minimal values for each queue length $q$, i.e., $TBs_q^{\max}$ or $TBs_q^{\min}$, respectively.
With the optimal blocklength being employed, the threshold-based optimal variable-length coding can be finally obtained by formulating the degenerated LP problem.
\vspace{-1mm}
\subsection{The Optimal Variable-Length Coding Policy}

We first present the optimal variable-length coding policy under the optimal solution $x^\ast_{q,s}$ of LP problem (\ref{D_P_LP})
Considering $x_{q,s}$ is defined as $f_{q,s}\pi_q$, the optimal variable-length coding $\boldsymbol{F}^\ast=\{f^\ast_{q,s}:0\le{}q\le{}Q,0\le{}s\le{}S\}$ is expressed as
\vspace{-2mm}
\begin{equation}\label{x_f}
f^\ast_{q,s}=
\begin{cases}
\frac{x^\ast_{q,s}}{\pi_q^\ast}           &{\rm{if~}}\pi_{q}^\ast>0\\
\mathds{1}_{\{s=s^{\max}_q\}}	&{\rm{if~}}\pi_{q}^\ast=0,
\end{cases}
\vspace{-2mm}
\end{equation}
where the optimal steady-state probability $\pi_q^\ast$ is given by $\pi_q^\ast=\sum_{s=s_q^{\min}}^{s_q^{\max}}x^\ast_{q,s}$.
As the steady-state probability, $\pi_q^\ast$ is specifically determined by the initial queue length with $\boldsymbol{F}^\ast$ given.
The initial queue length is particularly formulated according to the state classification of \mbox{the Markov chain that is generated by} $\boldsymbol{F}^\ast$.
In other words, the initial queue length is determined based on the constitution of the closed classes of recurrent states.
By this means, the specific initial queue lengths are necessary to obtain the optimal delay-power tradeoff.

Based on the properties in \emph{Theorem} \ref{th_curve}, we then show the same optimal delay-power tradeoff is obtained by the optimal variable-length coding with an arbitrary initial queue length. 
To this end, we only need to show the following theorem.
\vspace{-1mm}
\begin{theorem}\label{co_recurrent}
	The Markov chain under the optimal variable-length coding policy exists the only one recurrent closed class.
\end{theorem}
\vspace{-1mm}
\begin{IEEEproof}
We first focus on the extreme points on the optimal delay-power tradeoff curve, which can completely present the piecewise linear curve according to \emph{Theorem} \ref{th_curve}.
More specifically, the extreme points can only generated by the vertices of the polyhedron given by the constraints (\ref{LP_c_1})-(\ref{LP_c_4}).
Based on the method in \cite[Theorem 4.2]{altman1999constrained}, we can show the vertices are only generated by the deterministic coding policies, under which $s[n]$ is determined with the probability as $1$.
Meanwhile, the deterministic policies are also shown satisfying the theorem by leading the contradiction, which is formulated based on the definition of the vertices.

For the other points on the curve, the optimal variable-length coding policies are constructed as the convex combinations of the policies for two neighboring extreme points, which is shown in \cite[Lemma 1]{chen2017delaytcom}.
In this way, the only recurrent closed class is given for each optimal variable-length coding.
\end{IEEEproof}

With an arbitrary initial queue length, the optimal delay-power tradeoff curve can be always obtained for the URLLC user by using the optimal variable-length coding policies.

\vspace{-1mm}
\subsection{The Optimal Blocklength of the Queue-Aware Variable-Length Channel Coding}
\vspace{-1mm}

Then, the delay-optimal blocklength of the channel coding is presented for the queue lengths.
With the queue length $q[n]$ given, the blocklength of the optimal variable-length coding $\boldsymbol{F}^\ast$ is particularly selected as $TBs_{q[n]}^{\max}$ or $TBs_{q[n]}^{\min}$.
In other words, we have $s[n]$ can only be given by $s_{q[n]}^{\max}$ or $s_{q[n]}^{\min}$ under the optimal variable-length coding policy.
More specifically, the optimal blocklength of the variable-length coding can be presented as following proposition.
\begin{proposition}\label{th_rate}
	The optimal variable-length coding policy \mbox{$\boldsymbol{F}^\ast=\{f^\ast_{q,s}:0\le{}q\le{}Q,~0\le{}s\le{}S\}$ satisfies that}
	\begin{equation}\label{pro_max_min}
		\begin{cases}
			f^\ast_{q,s}\ge{}0&~{\rm{if~}}s=s_q^{\max}~{\rm{or}}~s_q^{\min},\\
			f^\ast_{q,s}=0    &~{\rm{otherwise}}.
		\end{cases}
	\end{equation} 
\end{proposition}

\mbox{Intuitively,} the minimal blocklength $TBs_{q[n]}^{\min}$ is employed to steady the queue length within the power constraint.
In the other hand, the maximal blocklength $TBs_{q[n]}^{\max}$ is employed to minimize the queue length, by which the \mbox{minimized average delay} is obtained with the power efficiency optimized according to \emph{Corollary} \ref{co_power_eff}.

Therefore, the optimal variable-length coding policy can be completely expressed based on $f^\ast_{q,s_q^{\max}}$ and $f^\ast_{q,s_q^{\min}}$ for all the queue lengths.
Meanwhile, the optimal delay-power tradeoff can be obtained by only considering the degenerated variable-length coding policies that satisfy Eq. (\ref{pro_max_min}).
In particular, the degenerated variable-length coding policy $\widebar{\boldsymbol{F}}$ can be presented by redefining the condition probabilities as
\begin{align}
f_q^{\max}&=\Pr\{s[n]=s_q^{\max}|q[n]=q\},\\
f_q^{\min}&=\Pr\{s[n]=s_q^{\min}|q[n]=q\},
\end{align}
i.e., $f_q^{\max}\!=\!f_{q,s_q^{\max}}$ and $f_q^{\min}\!=\!f_{q,s_q^{\min}}$. By this means, the degenerated variable-length coding policy $\widebar{\boldsymbol{F}}$ is given by  $\{({f}_q^{\max},{f}_q^{\min}):0\!\le\!{}q\!\le\!{}Q\}$.

With the degenerated variable-length coding policies, the optimal delay-power tradeoff can be formulated based on the method in Section III.
To this end, the steady-state probability $\pi_q$ is first attained by formulating the balance equations.
For the degenerated variable-length coding $\widebar{\boldsymbol{F}}$, the balance equations are presented in the following three cases.
\vspace{-2mm}
\begin{case}
	When $0\le{}q\le{}A-1$, we have
	\vspace{-1.5mm}
	\begin{equation}
	\sum_{i=0}^q\alpha{}\pi_{i}=\sum_{i=q+1}^{q+S}(1-\alpha)f_{i}^{\max}\pi_{i};
	\vspace{-1mm}
	\end{equation}
\end{case}
\vspace{-2mm}
\begin{case}
	When $A\le{}q\le{}Q-A-1$, we have
	\vspace{-1mm}
	\begin{equation}
	\sum_{i=q-A+1}^q\!\!\!\!\!\!\!\!\alpha{}f_{i}^{\min}\pi_{i}=\!\!\!\!\sum_{i=q+1}^{\min\{Q,q+S\}}\!\!\!\!\!\!\!\!\!(1-\alpha)f_{i}^{\max}\pi_{i}+\!\!\!\!\sum_{i=q+1}^{\min\{Q,q+S-A\}}\!\!\!\!\!\!\!\!\!\!\!\!\!\!\alpha{}f_{i}^{\max}\pi_{i};
	\vspace{-1mm}
	\end{equation}
\end{case}
\vspace{-2mm}
\begin{case}
	When $Q-A\le{}q\le{}Q$, we have
	\vspace{-1mm}
	\begin{equation}
	\sum_{i=q-A+1}^q\!\!\!\!\!\!\!\!\alpha{}f_{i}^{\min}\pi_{i}=\!\!\!\!\sum_{i=q+1}^{Q}\!\!\!(1-\alpha)\pi_{i}+\!\!\!\!\sum_{i=q+1}^{\min\{Q,q+S-A\}}\!\!\!\!\!\!\!\!\!\!\!\!\alpha{}f_{i}^{\max}\pi_{i}.
	\vspace{-1mm}
	\end{equation}
\end{case}

The optimal delay-power tradeoff can be then presented by formulating the degenerated LP problem with the degenerated coding policies only employed.
More specifically, we have
\begin{subequations}\label{D_P_den_LP}
	\begin{align}
	\min_{\{x_{q,s}\}}&
	\begin{array}{l}
	\!\!\frac{1}{A\alpha}\sum_{q=0}^{Q}q(x_{q}^{\max}+x_{q}^{\min})
	\end{array}\\
	~\rm{s.t.~}&\begin{array}{l}
	\!\!\sum_{q=0}^{Q}P(s_q^{\max})x_{q}^{\max}+P(s_q^{\min})x_{q}^{\min}
	\end{array}\le{}P_{\rm{th}}\\
	&\!\!\begin{array}{l}
		\sum_{i=0}^q\alpha{}(x_{i}^{\max}+x_{i}^{\min})=\sum_{i=q+1}^{q+S}(1-\alpha)x_{i}^{\max},
	\end{array}\notag\\
	&0\le{}q\le{}A-1\label{LP_d_c_1}\\
	&\!\!\begin{array}{l}
	\sum_{i=q-A+1}^q\alpha{}x_{i}^{\min}=\sum_{i=q+1}^{\min\{Q,q+S\}}(1-\alpha)x_{i}^{\max}
	\end{array}\notag\\
	&\!\!\begin{array}{l}
	+\sum_{i=q+1}^{\min\{Q,q+S-A\}}\alpha{}x_{i}^{\max},\,A\le{}q\le{}Q\!-\!A\!-\!1
	\end{array}\label{LP_d_c_2}\\
	&\!\!\begin{array}{l}
	\sum_{i=q-A+1}^q\alpha{}x_{i}^{\min}=\sum_{i=q+1}^{Q}(1-\alpha)(x_{i}^{\max}+x_{i}^{\min})
	\end{array}\notag\\
	&\!\!\begin{array}{l}
	+\sum_{i=q+1}^{\min\{Q,q+S-A\}}\alpha{}x_{i}^{\max},~Q-A\le{}q\le{}Q
	\end{array}\label{LP_d_c_3}\\
	&\begin{array}{l}
	\!\!\sum_{q=0}^{Q}x_{q}^{\max}+x_{q}^{\min}
	\end{array}=1\label{LP_d_c_4}\\
	&~~\!\!x_{q}^{\max}\ge{}0, x_{q}^{\min}\ge{}0,~~\forall~q,
	\end{align}%
\end{subequations}
where the three different cases of the transitions of $q[n]$ are presented in Eqs. (\ref{LP_d_c_1})-(\ref{LP_d_c_3}), respectively.

\subsection{The Threshold-based Optimal Variable-Length Coding}

Based on the degenerated LP problem, the optimal variable-length coding $\boldsymbol{F}^\ast=\{({f^\ast}_q^{\max},{f^\ast}_q^{\min}):0\le{}q\le{}Q\}$ is finally presented by using the threshold-based structure on the queue length.
With the optimal solution $({x^\ast}^{\max}_{q},{x^\ast}^{\min}_{q})$ of LP problem (\ref{D_P_den_LP}), we have
\vspace{-1mm}
\begin{equation}
	\begin{cases}
		{f^\ast}^{\max}_q=\frac{{x^\ast}^{\max}_{q}}{\pi^\ast_{q}},~{f^\ast}^{\min}_q=\frac{{x^\ast}^{\min}_{q}}{\pi^\ast_{q}} &{\rm{if~}}{\pi^\ast_{q}}>0\\
		{f^\ast}^{\max}_q=1,~~~~~~~{f^\ast}^{\min}_q=0 &{\rm{if~}}{\pi^\ast_{q}}=0,
	\end{cases}
	\vspace{-1mm}
\end{equation}
where $\pi^\ast_{q}$ is given by ${x^\ast}^{\max}_{q}\!+\!{x^\ast}^{\min}_{q}$.
With the optimal variable-length coding policy $\boldsymbol{F}^\ast$ formulated, the threshold-based structure on the queue length is presented as follows.%
\vspace{-1mm}
\begin{proposition}\label{th_th}
	With the threshold $q^\ast$ on queue length, the optimal variable-length coding policy $\boldsymbol{F}^\ast$ \mbox{is expressed as}
	\vspace{-1mm}
	\begin{equation}
		\begin{cases}
			{f^\ast}^{\min}_q=1&{\rm{if~}}q<q^\ast\\
			{f^\ast}^{\max}_q=1&{\rm{if~}}q>q^\ast\\
			{f^\ast}^{\min}_q+{f^\ast}^{\max}_q=1&{\rm{if~}}q=q^\ast.
		\end{cases}
		\vspace{-1mm}
	\end{equation}
\end{proposition}

In this way, the delay-optimal variable-length coding is immediately determined based on the order relation of the queue length $q[n]$ with the threshold $q^\ast$.
Based on the numerical results, the threshold-based structure is also maintained for the optimal variable-length coding policy when the function $P(s)$ satisfies the second property given in the \emph{Corollary} \ref{co_power_eff}, i.e., $\frac{P}{s}$ is decreasing with $s$.

\section{Numerical Results}

In this section, the numerical results are presented for the URLLC user to validate the optimal delay-power tradeoff and the threshold-based optimal variable-length coding policy.
The practical URLLC scenario is employed for the queue-aware variable-length coding, in which we have $A=1$, $Q=7$, and $S=3$.
Meanwhile, the RB spans $T=0.125$ms in the time domain and $B=1440$KHz in the frequency domain.
By setting $\epsilon=10^{-7}$, $L=256$bits, and $N_0=-150$dBm, the function $P(s)$ is formulated for the variable-length coding.
In particular, we have $P(0)=0~$W, $P(1)=2.59\times10^{-7}~$W, $P(2)=4.355\times10^{-7}~$W,
and $P(3)=6.038\times10^{-7}~$W, by which the sequence $\{P(s):s=0,\cdots,S\}$ is concave on $s$.

\begin{figure}[t]
	\centering
	\includegraphics[width=1\columnwidth,height=0.7\columnwidth]{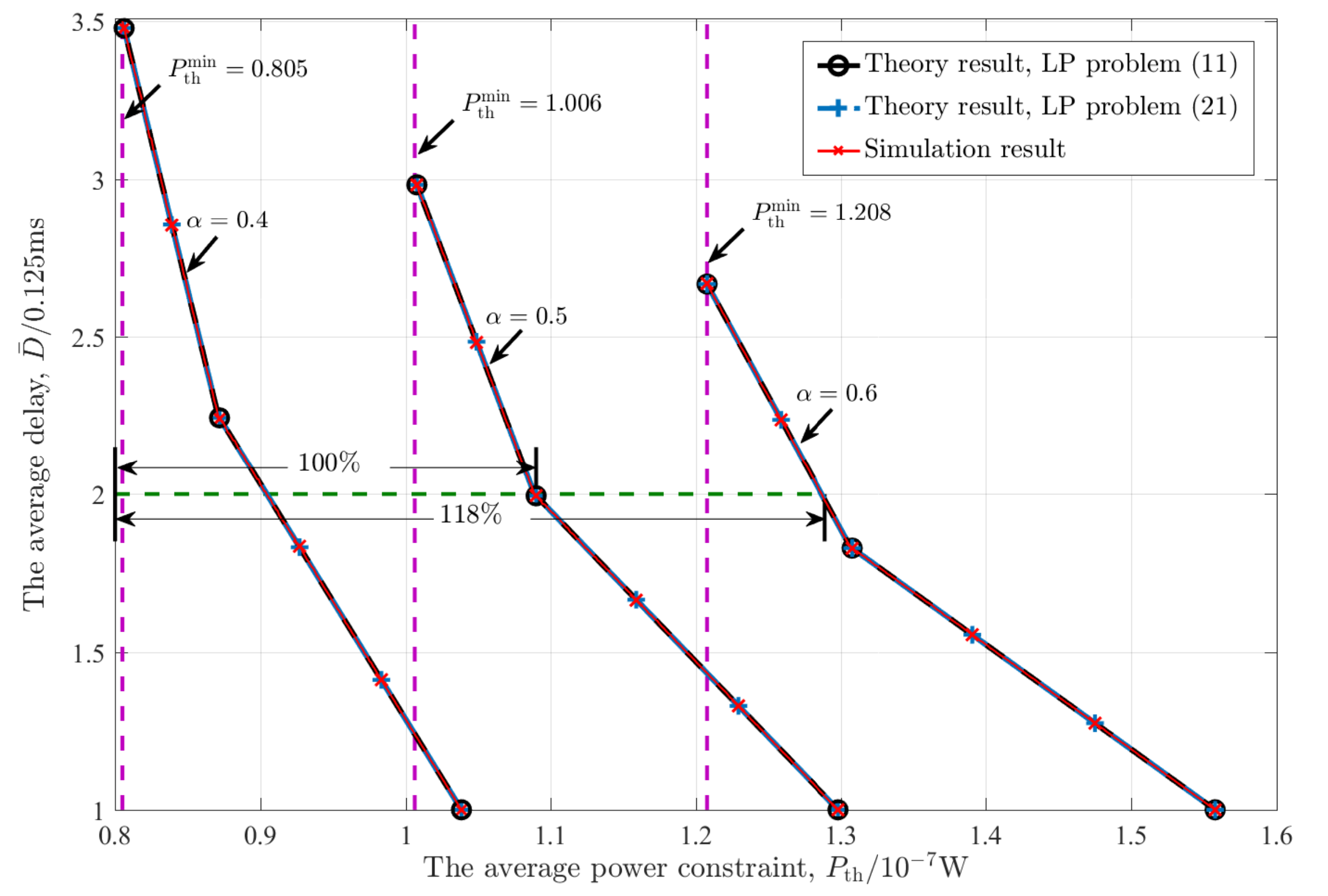}
	\vspace{-7mm}
	\caption{The numerical result of the optimal delay-power tradeoff.}
	\label{P_D_tradeoff}
	\vspace{-3mm}
\end{figure}

The optimal delay-power tradeoff curves are first given under the different arrival probabilities $\alpha$.
As demonstrated in Fig. \ref{P_D_tradeoff}, both of the theoretical results given by LP problems (\ref{D_P_LP}) and (\ref{D_P_den_LP}) well match with the Monte-Carlo simulation for the queue-aware variable-length coding.
More specifically, the optimal average delay will decrease with the increasing power constraint $P_{\rm{th}}$.
Meanwhile, the convexity and piecewise linearity for the optimal tradeoff curve can be also confirmed in Fig. \ref{P_D_tradeoff}.
To obtain the same average delay, the average power consumption for the optimal variable-length coding will increases with probability $\alpha$.
When the average delay is given by $0.25$~ms, the average power consumption with $\alpha$ as $0.6$ is about $118\%$ of that with $\alpha$ as $0.5$.
To steady the queue length, the minimal power constraint $P_{\rm{th}}^{\min}$ is presented for the each $\alpha$.
With the arrival probability $\alpha$ increasing, the minimal power constraint $P_{\rm{th}}^{\min}$ is also increasing.

Then, we present the specific optimal variable-length coding policy, in which the threshold-based structure is given on the queue length.
In Table \ref{typical_policy}, the optimal variable-length coding policy $\boldsymbol{F}^\ast$ is presented by using the condition probabilities $f^\ast_{q,s}$.
Meanwhile, the probabilities $f^\ast_{q,s}$ for the pairs $(q,s)$ that induce to the overflow or underflow are shown in gray.
In each column of Table \ref{typical_policy}, the probabilities $f_{q,s}^\ast$ under the same queue length are presented.
In particularly, the $f_{q,s}^\ast$ are equal to zero except $f_{q,s^{\max}_q}^\ast$ and $f_{q,s^{\min}_q}^\ast$, i.e., $f_{q}^{\max}$ and $f_{q}^{\min}$.
By this means, the blocklength of the variable-length coding is only selected as $TBs_{q[n]}^{\max}$ or $TBs_{q[n]}^{\min}$.
Considering the threshold $q^\ast$ on the queue length is equal to $2$, the optimal variable-length coding policy is then formulated according to \emph{Proposition} \ref{th_th}.
The delay-optimal blocklength is particularly given by $TBs_{q[n]}^{\min}$ or $TBs_{q[n]}^{\max}$ when the queue length $q[n]$ is less or greater than the threshold $q^\ast$, respectively.
When $q[n]$ is equal to $q^\ast$, both of the maximal and minimal blocklengths will be employed with the probabilities as $0.5$.

\vspace{-1mm}
\section{Conclusions}

In this paper, the queue-aware variable-length coding has been studied in URLLC based on the cross-layer approach, where the blocklength of coding is determined by the queue length.
In this way, the delay-optimal variable-length coding has been attained in the single user link by formulating the optimal delay-power tradeoff for the URLLC user.
Meanwhile, the reliability requirement in URLLC has been also satisfied by the power allocation of the variable-length coding.
By presenting the equivalent LP problem, the minimized average delay has been particularly presented for the URLLC user under the average power constraint.
Moreover, the optimal variable-length coding has been presented based on the threshold-based structure on the queue length, under which the delay-optimal blocklength is determined for the variable-length coding.

\begin{table}[t]
	\caption{The threshold-based optimal variable-length coding policy }
	\vspace{-2mm}
	\footnotesize
	\label{typical_policy}
	\centering
	\setstretch{1}
	\begin{tabular}{|c|c|p{0.2in}<{\centering}|p{0.2in}<{\centering}<{\arrayrulecolor{red}}|p{0.2in}<{\centering}<{\arrayrulecolor{red}}|p{0.2in}<{\centering}<{\arrayrulecolor{black}}|p{0.2in}<{\centering}|p{0.2in}<{\centering}|p{0.2in}<{\centering}|p{0.2in}<{\centering}|}
		\hline
		\multicolumn{2}{|c|}{\multirow{2}{*}{$f_{q,s}^{\ast}$}}&\multicolumn{8}{c|}{$q[n]$}\\
		[1pt] 
		\cline{3-10}
		\multicolumn{2}{|c|}{~}&0&1&\multicolumn{1}{c|}{\red{2}}&3&4&5&6&7\\
		\hline
		\multirow{4}{*}{$s[n]$}
		&0&1&1&\multicolumn{1}{c|}{0.5}&0&0&0&0&\gray{0}\\
		\cline{2-10}
		&1&\gray{0}&0&\multicolumn{1}{c|}{0}&0&0&0&0&0\\
		\cline{2-10}
		&2&\gray{0}&\gray{0}&\multicolumn{1}{c|}{0.5}&0&0&0&0&0\\
		\cline{2-10}
		&3&\gray{0}&\gray{0}&\multicolumn{1}{c|}{\gray{0}}&1&1&1&1&1\\
		\hline
	\end{tabular}
\end{table}


\footnotesize
\bibliographystyle{IEEEtran}
\bibliography{ref_scheduling}
\normalsize
\end{document}